\documentclass{mem}
\usepackage{natbib}
\usepackage{txfonts}
\usepackage{balance}
\usepackage{graphicx}
\usepackage[a4paper]{hyperref}
\begin{document}

\title{
Cloud modeling of a quiet solar region in H$\alpha$
}

\subtitle{}

\author{
Z. F. \,Bostanc\i\inst{1,2} 
\and N. \,Al Erdo\u{g}an\inst{1} 
}

\offprints{Z. F. Bostanc\i}

\institute{
Istanbul University, University Observatory, 34119 University-Istanbul, Turkey
\and 
Sterrekundig Instituut Utrecht, Postbus 80 000, 
NL–3508 TA Utrecht, The Netherlands
\email{fbostanci@gmail.com}
}

\authorrunning{Bostanc\i\ \& Al Erdo\u{g}an }

\titlerunning{Cloud modeling of a quiet solar region in H$\alpha$}

\abstract{
We present chromospheric cloud modeling on the basis of 
H$\alpha$ profile-sampling images taken with the Interferometric 
Bidimensional Spectrometer (IBIS) at the Dunn Solar Telescope (DST). 
We choose the required reference background profile by using 
theoretical NLTE profile synthesis.  The resulting cloud parameters 
are converted into estimates of physical parameters (temperature and 
various densities). Their mean values compare well with the VAL-C model.  
}

\maketitle{}

\section{Introduction}

The solar chromosphere observed in H$\alpha$ shows a mass of fibrilar
structures. They are called mottles when seen on the disk, spicules
when seen at the limb. Studies of these structures are important to
understand chromospheric dynamics and its contribution to
outer-atmosphere heating.

Chromospheric observations sampling spectral profiles give the
opportunity to derive physical properties per fine structure. We do
that here for H$\alpha$ using the DST/IBIS data of \cite{cauzzi09}.
Cloud modeling following \citet{beckers64} is the usual method to
obtain line formation parameters by matching the observed contrast
profile of a structure with a theoretical one \citep[see review
by][]{tziotziou07}. This approach can only be used if the studied
structure is fully separated from the underlying atmosphere, and
necessitates description of the latter by a background profile. Its
choice or determination is crucial \citep[see e.g.][]{durrant75}. We
address this issue by trying different synthetic H$\alpha$ profiles.
We then use the method of \citet{tsiropoula97} to derive physical
parameters from the cloud model results.

\section{Observations}

\begin{figure}[t!]
\center
\resizebox{0.9\hsize}{!}{
\includegraphics[clip=true]{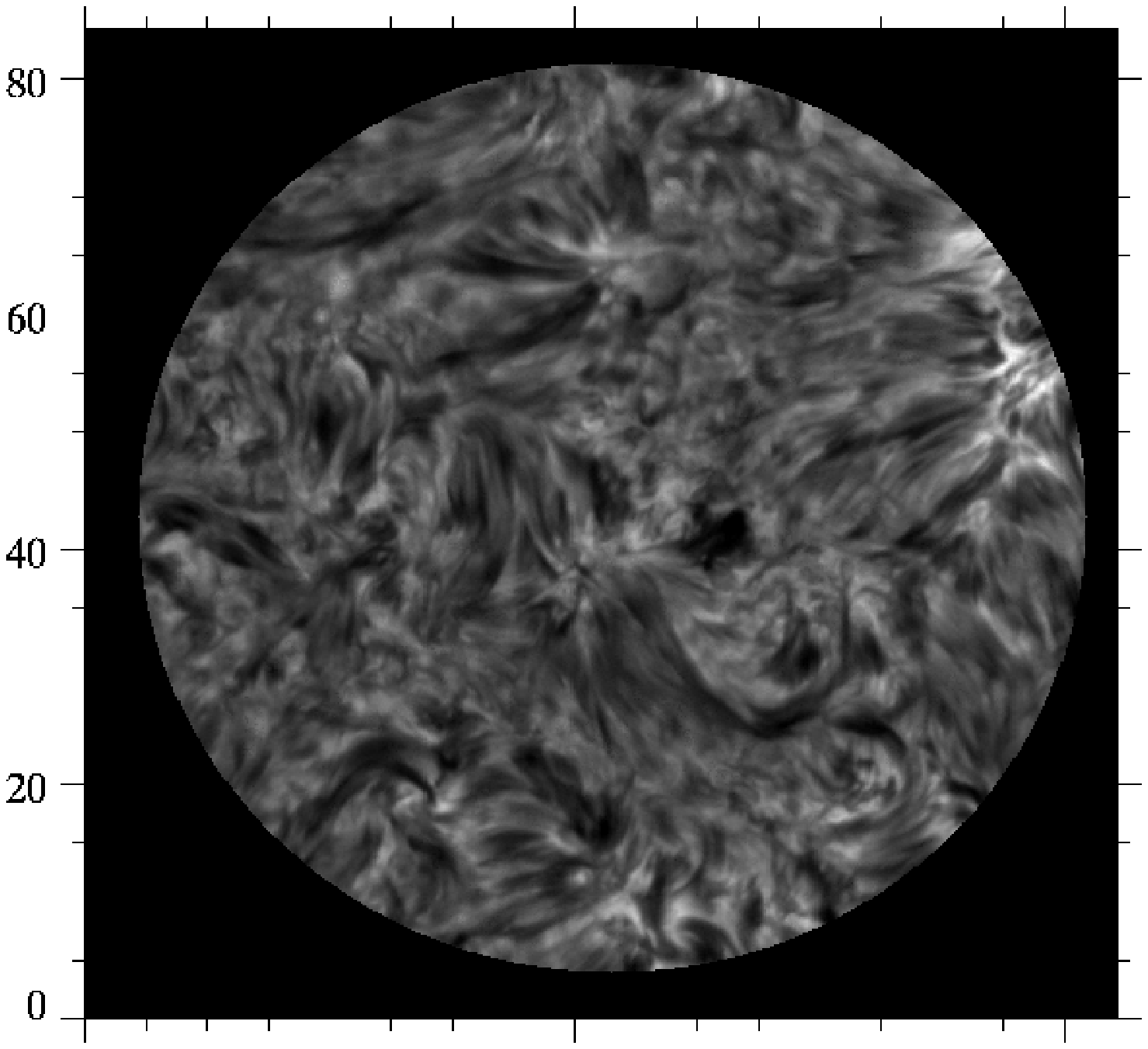}}
\resizebox{0.9\hsize}{!}{
\includegraphics[clip=true]{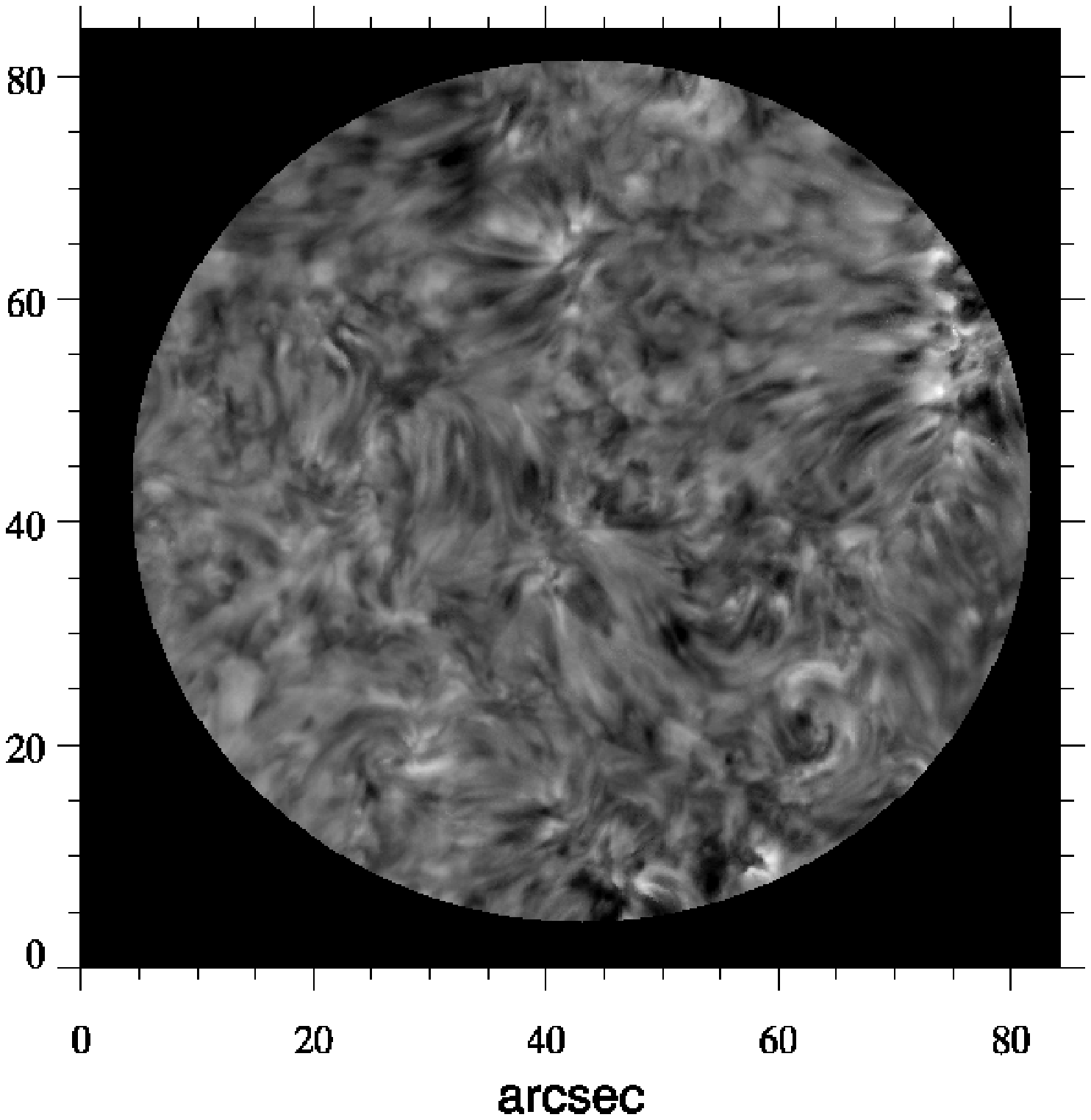}}
\caption{\footnotesize 
\emph{Upper image}: profile-minimum intensity. 
\emph{Lower image}: profile-minimum Dopplershift, 
with blueshift black, redshift white. }
\label{fig:centerimg}
\end{figure}

In March 2007 a quiet-sun area near disk center was observed in
H$\alpha$ with IBIS at the DST \citep{cavallini06,reardon08}. These
observations were presented and analyzed by \cite{cauzzi09}. The
H$\alpha$ line was sampled at 24 spectral positions at step intervals
of 90~m\AA in H$\alpha$ in a sequence of 192 spectral scans at a
cadence of 15.4~seconds. Line profiles were constructed for each pixel
in the field of view (diameter 80 arcsec). For each spectral profile,
the line-center wavelength was established by fitting a polynomial to
the five spectral samplings with least intensity. The minimum of the
fit defines the per-pixel intensity minimum and line-of-sight
velocity. Figure~\ref{fig:centerimg} shows the minimum intensity and
Doppler velocity images from a single scan taken at one of the best
seeing moments. We refer to \cite{cauzzi09} for more detail.
%

\section{Results \& Conclusions}

\subsection{Cloud Model}

The traditional cloud model delivers the four parameters source
function $S$, optical thickness at line center $\tau_{0}$, Doppler
width $\Delta\lambda_{\mathrm{D}}$, and line of sight velocity
$\upsilon_{\mathrm{LOS}}$. The model assumes an optically thin, homogeneous
cloud that is illuminated by uniform radiation from below, so that
these parameters are assumed constant along the line of sight through
the cloud. The observed contrast profiles are then matched with
theoretical contrast profiles given by: 
\begin{equation}
  {{I(\lambda)-I_0(\lambda)}\over{I_0(\lambda)}}
  =\left({S\over{I_0(\lambda)}}-1\right)
   \left(1-e^{{-}\tau(\lambda)}\right),
\label{eq:equation1}
\end{equation}
where $I(\lambda)$ is the local profile, $I_0(\lambda)$ the reference
background profile and $\tau(\lambda)$ the optical thickness
\begin{equation}
  \tau(\lambda)=\tau_0\exp\left[-\left(
     {\lambda-\lambda_c(1-\upsilon_{\rm{LOS}}/c)
     \over \Delta\lambda_D}\right)^2\right].
\label{eq:equation2}
\end{equation}

The parameter fitting is achieved by iterative least-square matching
of the observed contrast profile with a theoretical one.
\begin{figure}
\center
\resizebox{\hsize}{!}{
\includegraphics[clip=true]{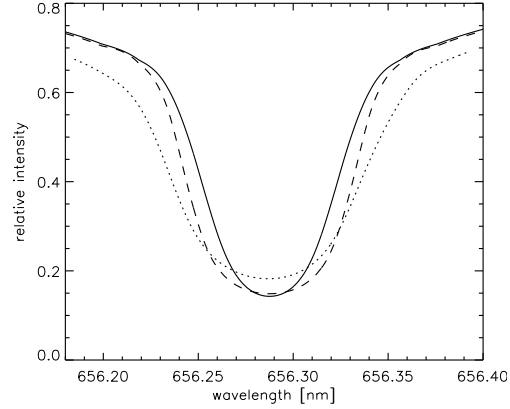}}
\caption{\footnotesize 
Background profiles, with the intensities normalized to the
continuum value. \emph{Solid}: Kurucz model. \emph{Dashed}: 
FAL-C model. \emph{Dotted}: observed mean profile.}
\label{fig:background}
\end{figure}
\begin{figure}
\center
\resizebox{\hsize}{!}{
\includegraphics[clip=true]{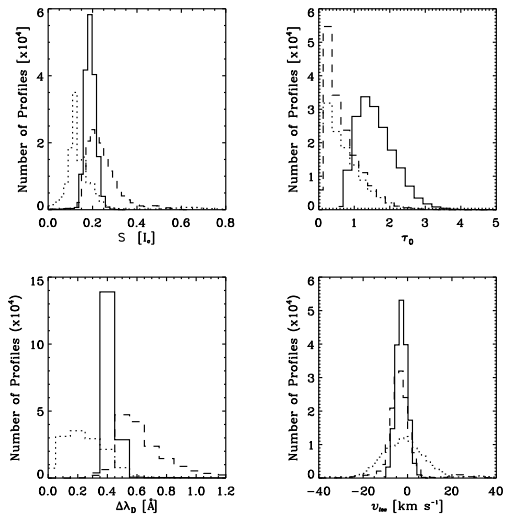}}
\caption{\footnotesize
Occurrence distributions of the cloud parameters. 
Clockwise source function $S$, optical thickness at line center 
$\tau_{0}$, Doppler width $\Delta\lambda_{\rm D}$, and line of sight 
velocity $\upsilon_{\mathrm{LOS}}$. \emph{Solid}: Kurucz background 
profile. \emph{Dashed}: FAL-C background profile. \emph{Dotted}: 
observed mean profile used as background profile.  }
\label{fig:distbgprof}
\end{figure}
%
\begin{figure}[t!]
\resizebox{\hsize}{!}{
\includegraphics[clip=true]{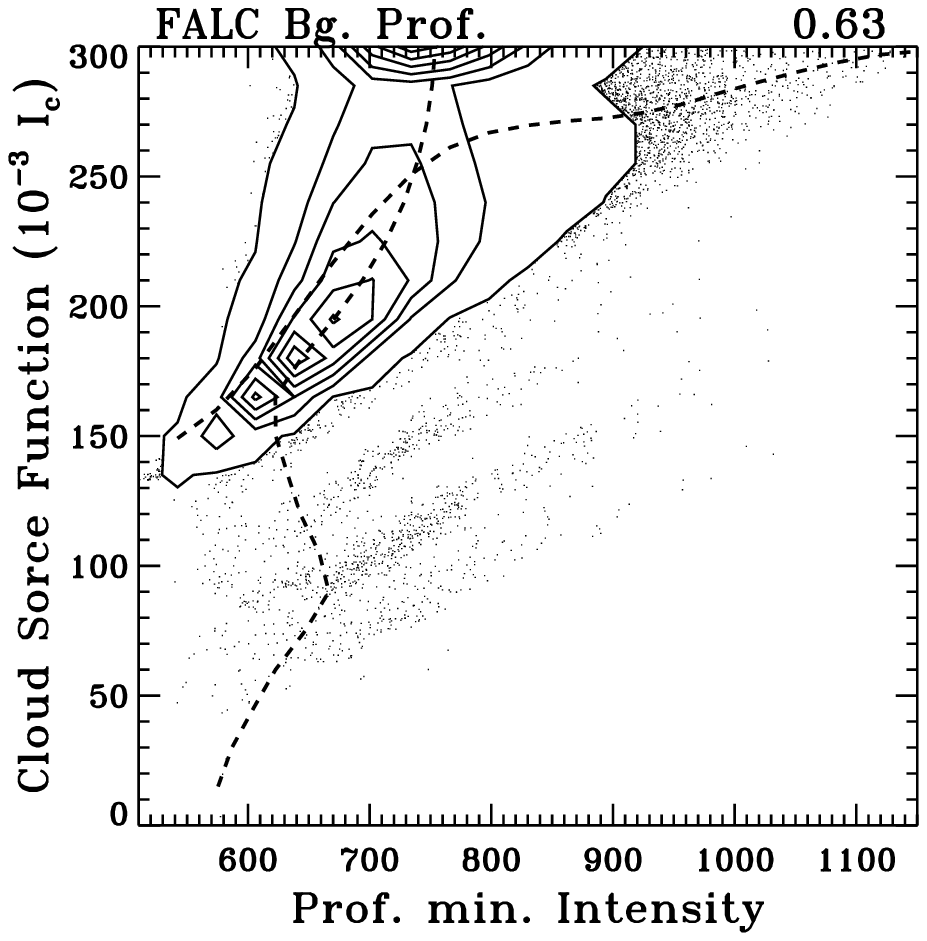}
\includegraphics[clip=true]{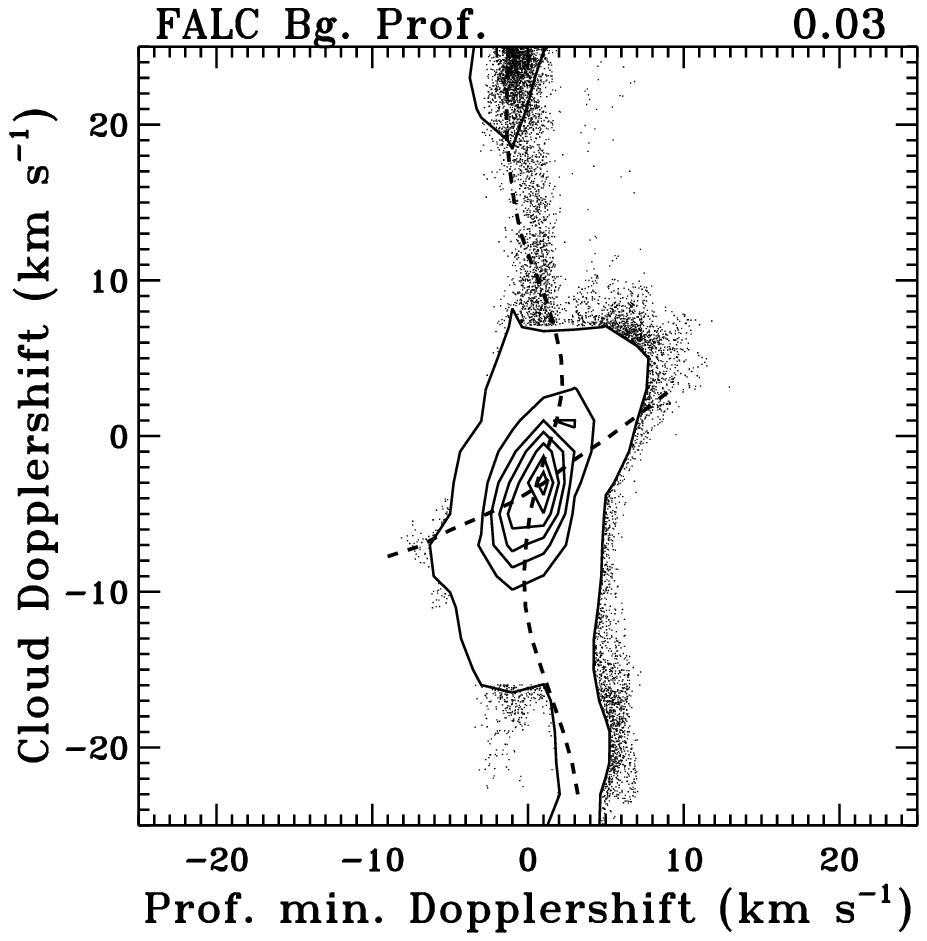}}
\resizebox{\hsize}{!}{
\includegraphics[clip=true]{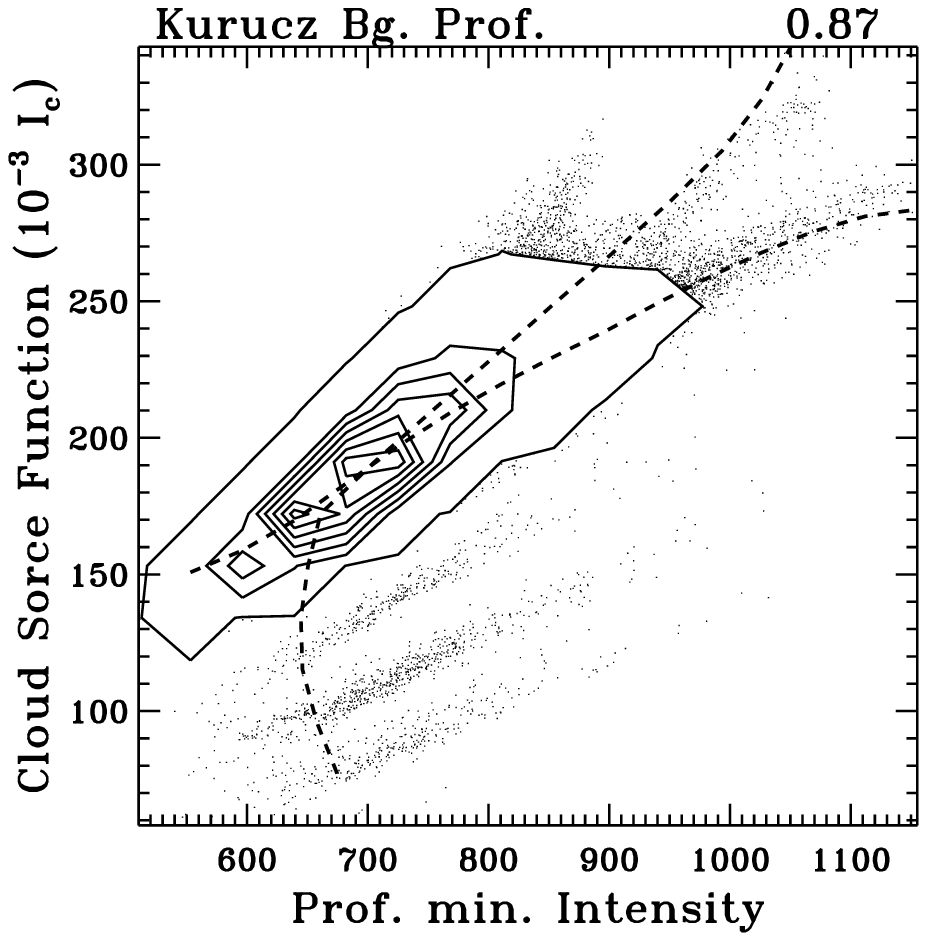}
\includegraphics[clip=true]{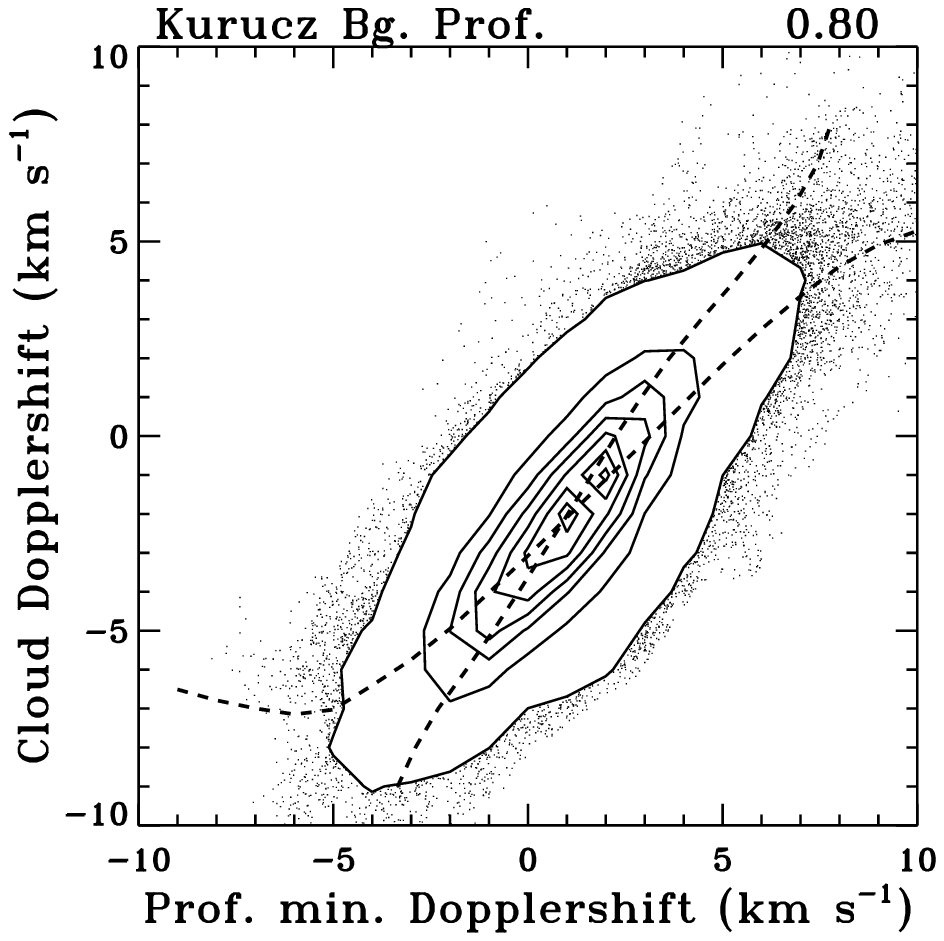}}
\resizebox{\hsize}{!}{
\includegraphics[clip=true]{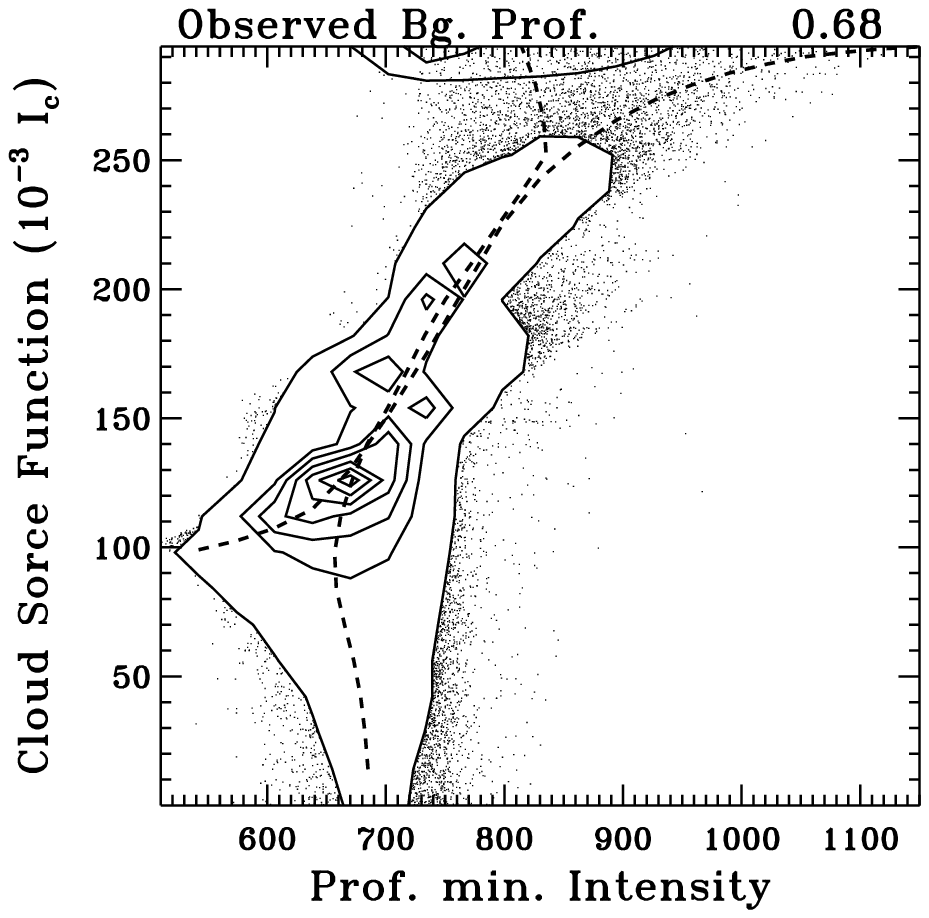}
\includegraphics[clip=true]{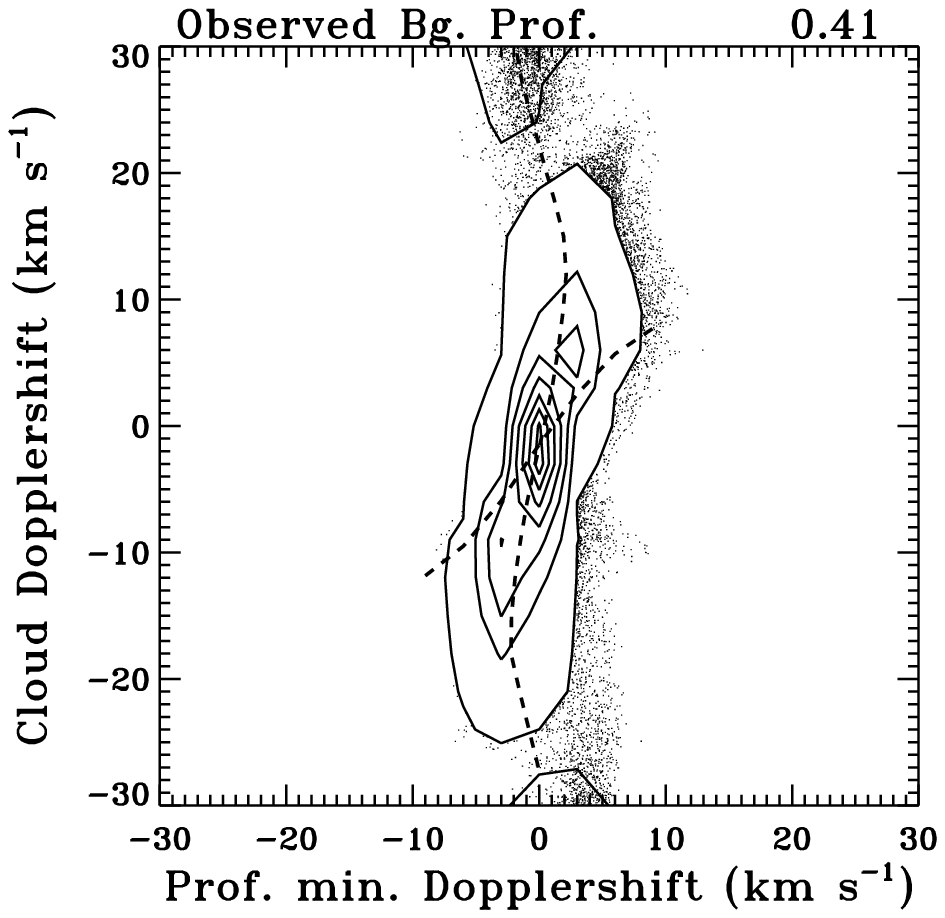}}
\caption{\footnotesize
Scatter plots. \emph{Left column}: H$\alpha$ source 
function $S$ against the observed profile-minimum intensity, 
respectively with the FAL-C background profile (top), the Kurucz 
background profile (middle), and the observed mean profile as 
background profile (bottom). \emph{Right column}: idem for the 
fitted line-of-sight cloud velocity $\upsilon_{\rm LOS}$ against the 
observed profile-minimum Dopplershift. The numbers at the 
upper-right corners specify the overall Pearson correlation coefficient. }
\label{fig:plotscat}
\end{figure}

\subsection{Background profile}

The background profile I$_0(\lambda)$ represents the irradiation from 
supposedly plane-parallel atmosphere underlying the cloud-like 
chromospheric structure. Its choice has a significant effect  on the
resulting cloud parameters. We here present H$\alpha$ cloud modeling
results for the IBIS scan with the best seeing using three different
background profiles: a synthetic profile computed with a 
one-dimensional NLTE line formation code from the FAL-C
\citep{fontenla93} standard model which contains a chromospheric
temperature rise, a synthetic NLTE profile similarly computed from the
Kurucz \citep{kurucz79,kurucz92a,kurucz92b} radiative-equilibrium
model in which the temperature declines outward without chromosphere,
and the spatial-temporal average of all observed H$\alpha$ profiles
over the full field of view during the whole 50-min time series. The
three profiles are shown in Figure~\ref{fig:background}.

With each  background profile, cloud-model fitting was applied to the
observed  H$\alpha$ profile at all $1.74\times10^5$  pixels in the
IBIS field of view. We rejected  the pixels with resulting values $S >
0.8 I_{\mathrm{c}}$ (in units of  continuum intensity) and also the pixels giving
$\tau_0 > 5$. In addition, the cloud model routine did not converge or
did not yield physically acceptable values for 0.04\%, 7.47\% and
23.19$\%$ of  the total  when we used  the Kurucz, FAL-C, and observed
mean  profile, respectively. Figure~\ref{fig:distbgprof} displays the
remaining  distributions of the cloud model parameters for each
background profile. It shows that using the Kurucz profile permits a
solution for many more H$\alpha$ profiles, with narrower parameter
distributions.

Figure~\ref{fig:plotscat} shows scatter plots for the resulting values
of $S$ and $\upsilon_{\mathrm{LOS}}$ against the observed profile-minimum
intensity and Dopplershift, respectively, when using the three
different background profiles. The best correlations are found between
these cloud parameters and profile-minimum measurements when the
Kurucz synthetic profile is used.

These comparisons suggest that the synthetic Kurucz profile is the
best choice as background profile for cloud modeling of these
H$\alpha$ observations. The spread between these tests confirms that
the issue of the selection of a background profile is a key one in
chromospheric cloud modeling.

\subsection{Physical parameters} 

We then applied cloud model fitting using the Kurucz background
profile to all 192 spectral scans in the 50-min time series. Afterwards, we
converted the resulting cloud parameters into more physical parameters
with the method of \citet{tsiropoula97}. The resulting mean values and
standard deviations are given in Table~\ref{tbl:obsvstheo}. For
comparison the values of the VAL-C atmosphere model of
\citep{vernazza81} at the height where its $N_2$ population is close
to the mean value in the cloud determinations are also listed.
Table~\ref{tbl:obsvstheo} shows good agreement between the cloud model
and VAL-C values.
%
\begin{table}
\centering
\caption[]{\label{tbl:obsvstheo} 
Comparison of the observed parameters with the values inferred from VAL-C atmosphere models.}
\setlength{\tabcolsep}{4pt}
\begin{tabular}{lcc}
\hline
Physical & \multicolumn{2}{c} {Quiet Chromosphere}\\
Parameters&Observational&VAL-C\\
\hline
$S$~($I_{\mathrm{c}}$)            & $0.19 \pm 0.02$ & $-$\\
$\tau_{0}$                       & $1.50 \pm 0.46$ & $-$\\
$\Delta\lambda_{\mathrm{D}}$~(\AA)& $0.46 \pm 0.04$ & $-$\\
$\upsilon_{\mathrm{LOS}}$~(km~s$^{-1}$) & $-1.74 \pm 3.37$ & $-$ \\
$N_{1}$~($10^{10}$~cm$^{-3}$)     & $2.21 \pm 0.42$ & 1.24\\
$N_{2}$~($10^{4}$~cm$^{-3}$)      & $2.54 \pm 0.88$ & 2.88\\
$N_{\mathrm{e}}$~($10^{10}$~cm$^{-3}$) & $5.13 \pm 0.94$ & 3.54\\
$N_{\mathrm{H}}$~($10^{10}$~cm$^{-3}$) & $8.02 \pm 1.47$ & 4.67\\
$T$ ($10^{4}$~K)                 & $1.43 \pm 0.40$ & 1.07\\
$P$~(dyn~cm$^{-2})$              & $0.27 \pm 0.09$ & 0.13\\
$M$~($10^{-5}$~gr~cm$^{-2}$)      & $3.61 \pm 0.55$ & 0.62\\
$\rho$~($10^{-13}$~gr~cm$^{-3}$)  & $1.75 \pm 0.28$ & 1.09\\
$\chi_{\mathrm{H}}$               & $0.63 \pm 0.01$ & $-$\\
\hline
\end{tabular}
\end{table}


\begin{acknowledgements} 
We are indebted to Kevin Reardon for supplying the observations and 
explanation, to Han Uitenbroek for suggesting and computing the 
theoretical background profiles, to Rob Rutten for improvement of 
this paper, and to all three and Gianna Cauzzi and Alexandra 
Tritschler for discussions.  This research project has been 
supported by a USO--SP Marie Curie Early Stage Research Training 
Fellowship from the European Community under contract number 
MEST-CT-2005-020395.  IBIS was built by INAF/Osservatorio 
Astrofisico di Arcetri with contributions from the Universita di 
Firenze and the Universita di Roma Tor Vergata. IBIS construction 
and operation has been supported by the Italian Ministero 
dell'Universita e della Ricerca (MUR), as well as the Italian 
Ministry of Foreign Affairs (MAE). NSO is operated by the 
Association of Universities for Research in Astronomy, Inc. (AURA), 
under cooperative agreement with the National Science 
Foundation. This work was supported by the Research Fund of Istanbul 
University as project number 851.
\end{acknowledgements}

\bibliographystyle{aa}

\end{document}